\documentclass[nofootinbib,twocolumn,
showpacs,preprintnumbers,amsmath,amssymb]{revtex4}
\usepackage{latexsym}
\usepackage{hyperref}
\usepackage{epsfig, psfrag, graphicx}
\usepackage{bm,dsfont}

\newcommand{\be}{\begin{equation}}
\newcommand{\ee}{\end{equation}}
\newcommand{\ben}{\begin{eqnarray*}}
\newcommand{\een}{\end{eqnarray*}}
\newcommand{\bea}{\begin{eqnarray}}
\newcommand{\eea}{\end{eqnarray}}
\newcommand{\bdm}{\begin{displaymath}}
\newcommand{\edm}{\end{displaymath}}
\newcommand{\ba}{\begin{align}}
\newcommand{\ea}{\end{align}}
\newcommand{\lb}{\label}
\renewcommand{\d}{\operatorname{d}\!}
\renewcommand{\exp}{\operatorname{exp}\!}
\renewcommand{\cosh}{\operatorname{cosh}\!}
\renewcommand{\sinh}{\operatorname{sinh}\!}
\renewcommand{\cos}{\operatorname{cos}\!}
\renewcommand{\sin}{\operatorname{sin}\!}

\renewcommand{\ln}{\operatorname{ln}\!}

\begin{document}

\title{\bf Quantum Cosmology Close to the Classical Big Bang Singularity and in the Semiclassical Limit}
\author{Frank Steiner\footnote{e-mail address:{\tt frank.steiner@uni-ulm.de}} and Andreas J. W\"ohr\footnote{e-mail address:{\tt andreas.woehr@uni-ulm.de}}}
\affiliation{Institut f\"ur Theoretische Physik, Universit\"{a}t Ulm, Albert-Einstein-Allee 11,
89069 Ulm, Germany}

\begin{abstract}
We investigate a cosmological model whose energy content is described by a Chaplygin gas represented by a scalar field $\phi$ with an associated potential producing a big bang singularity such that for vanishing scale factor, $a\rightarrow 0$, one has $\left|\phi\right|\rightarrow \infty$. The classical version of the model is discussed in detail, however, our main interest lies in its quantization. Upon quantization of this model in the Schr\"odinger picture, we get the Wheeler--DeWitt equation which can be solved exactly in the two limits $a\rightarrow 0$ and $a\rightarrow\infty$, respectively. Employing the DeWitt criterium that the wave function should vanish at the classical singularity in order to avoid the big bang, we show that a solution to the Wheeler-DeWitt equation fulfilling this condition can indeed be found. In addition to DeWitt's initial condition at the big bang, we postulate an asymptotic condition to be imposed on the wave function which guarantees that the quantum wave function is strongly peaked at the classical field configurations in the semiclassical limit. We also investigate a universe filled with dust (describing baryonic and dark matter) and show that in this case there exists an exact solution to the Wheeler-DeWitt equation which describes the evolution of the universe during its whole history. Also for this model one can construct a wave packet which avoids the big bang singularity at small scale factors and is strongly peaked at the classical field in the far future of the universe. Finally, we discuss the significance of our treatment of the Wheeler-DeWitt equation and mention some open problems.     
\end{abstract}

\pacs{04.60.Ds, 
      98.80.Qc  
              }

\maketitle

\section{Introduction}
The absence of singularities may be taken to be a prerequisite for any fundamental theory of nature, and in particular we expect a quantum theory of gravity to resolve the curvature singularities found in Einstein's theory of general relativity. However, up to now there exists no complete quantum theory of gravity. Because of that, the problem of quantum cosmology has been approached by first carrying out a symmetry reduction (by requiring isotropy and homogeneity) and then quantizing the resulting minisuperspace models which have only a finite number of degrees of freedom \cite{DeWitt_67, Misner_69, C.Kiefer_07}. 
One of the main approaches to quantum cosmology is based on a canonical quantization of general relativity. In that approach, there exist basically two promising candidates for a quantum cosmological theory (see e.g. the textbooks \cite{C.Kiefer_07} and \cite{Thiemann_07}): on the one hand, minisuperspace quantization in the framework of the geometrodynamical approach, and on the other hand loop quantum cosmology (LQC hereafter) \cite{BojowaldR}.

In the geometrodynamical as well as in the LQC approach, one of the important problems is to investigate whether the classical singularities can be avoided. This includes that for each approach one has to define what the expression "singularity avoidance" means. Both approaches describe the universe via a
wave function on configuration space which has to be the solution of constraint equations. In a canonical setting, the dynamics is implemented by the Hamiltonian constraint. The difference between both approaches lies in the way this equation is quantized. LQC is the quantization of homogenous and isotropic minisuperspaces, where one uses a so-called polymer representation instead of the conventional Schr\"odinger representation based on the full theory of loop quantum gravity (LQG hereafter) \cite{Ashtekar_04,Thiemann_03,Rovelli_04}. As in LQG, the underlying geometry in LQC is discrete and the scale factor operator has discrete eigenvalues. Quantum dynamics is governed by a discrete difference equation in eigenvalues $V_{\mu}$ of the volume operator $\hat V$ \cite{Ashtekar_06*,Ashtekar_07}. In the Schr\"odinger picture, one arrives at a differential equation, the Wheeler--DeWitt equation \cite{Wheeler:64,DeWitt_67}. In the continuum limit for large volume, the Wheeler--DeWitt equation is recovered from the loop quantum cosmological difference equation \cite{Ashtekar_06}.  

In this paper, we restrict the discussion to minisuperspace quantization in the Schr\"odinger picture. In the Schr\"odinger picture, several models have been investigated regarding their ability to cure the singularity problem which is prevalent in the classical theory of general relativity. In this setting, singularity avoidance is defined as either a vanishing of the wave function at the point of the classical singularity (DeWitt criterium) or a spreading of semiclassical states indicating a break-down of semiclassical concepts in general. In the following, we analyze whether the big bang singularity can be avoided and how reliable those solutions of the Wheeler-DeWitt equation are.

We first discuss at length the Chaplygin gas \cite{Chap_04} which presents one of the simplest models possessing a big bang singularity. The Chaplygin gas is a perfect fluid which satisfies the equation of state $p = -A/\epsilon$, where $A$ is a positive constant. The Chaplygin gas has acquired some popularity in cosmology as candidate for unifying dark sectors of the universe, that is dark energy and dark matter, introduced by Kamenshchik et al. \cite{KAM_01}, and has been widely studied since then in this context (see, e.g., \cite{Fabris_02,Fabris_02*,Gorini_03,Gorini_05,Chimento_03,Chimento_04,Dev_03,Avelino_03,Bean_03,Carturan_03}). Furthermore, in particle physics the Chaplygin gas has also raised interest due to its connection with string theory \cite{Bordemann_93} and its supersymmetric extension \cite{Jackiw_00}. The problem of singularity avoidance has also been investigated for an anti-Chaplygin gas for which the constant $A$ above is negative \cite{KAM_07}.

We also investigate a pure dust model starting with a big bang which describes baryonic and dark matter and plays the role of a prototype model for our Universe.

This paper is organized as follows. In  the next Section we briefly review the minisuperspace model.
In Sec.~\ref{klass} we discuss in detail the Chaplygin gas as a simple classical model exhibiting a big bang singularity. In Sec.~\ref{sec:quant} the Wheeler-DeWitt equation for this model is studied and the exact solutions for the quantum wave functions are given for the two physically interesting limiting cases. In Sec.~\ref{sec:bb} the Wheeler-DeWitt equation is solved in the vicinity of the classical big bang singularity, and in Sec.~\ref{sec:largevalues} in the limit where the cosmic scale factor goes to infinity, that is for the far future of the universe. It is shown in Sec.~\ref{BB} that one can construct in a natural way solutions to the Wheeler-DeWitt equation describing wave packets which vanish at the classical big bang and thus fulfill DeWitt's criterium. In Sec.~\ref{semisteiner} we present a detailed discussion of the wave function of the Chaplygin gas in the semiclassical limit using the method of stationary phase. We explicitly show how a wave packet can be constructed which is for large scale factors in leading approximation strongly peaked exactly at the classical field trajectories. Sec.~\ref{dust} is devoted to the treatment of the dust model for which the Wheeler-DeWitt equation is solved exactly in the whole minisuperspace. Again we are able to construct a wave packet which avoids the classical big bang. Finally, Sec.~\ref{discussion} contains a summary and discussion. 

\section{the minisuperspace model}
\lb{mini}
In most investigations, the problem of applying quantum gravity to cosmology is simplified by a symmetry reduction to homogeneous and isotropic geometries. In this Section, the symmetry reduction is performed on the classical level leaving only one gravitational degree of freedom given by the cosmic scale factor $a(t)$. Restricting to a flat Friedmann-Lema\^\i tre universe, the metric is given by the Robertson-Walker metric ($c=1$)
\be
\lb{line element}
ds^2 = dt^2-a^2(t)d\vec{x}^2\ ,
\ee
where we have set the lapse function equal to $N=1$. In order to take the energy content of the universe into account, we consider the universe to be filled with a perfect fluid with energy density $\epsilon$ and pressure $p=p(\epsilon)$. At the quantum level, we need a more fundamental description. Thus, in the following, we mimic the perfect fluid by a homogeneous scalar field $\phi(t)$ in terms of which $\epsilon$ and $p$ are given by
\be
\lb{ep}
\epsilon=\frac{\dot\phi^2}2+V(\phi), \quad p=\frac{\dot\phi^2}2-V(\phi)  \ .
\ee
Here $V(\phi)$ is the potential of the scalar field $\phi$. In Sec.~\ref{klass} we shall derive an explicit expression for  $V(\phi)$ by demanding that the perfect fluid is a Chaplygin gas, and in Sec.~\ref{dust} for a pure dust model.

The full action, being the sum of the Einstein-Hilbert action and the action of the minimally coupled scalar field, is then given by
\bea
\lb{action}
S[a,\phi] =  \frac{3\mathcal{V}_0}{\kappa} \int dt  \left(-a\dot{a}^2\right) + \mathcal{V}_0 \int dt a^3\left[\frac{\dot{\phi}^2}{2}-V(\phi)\right]\nonumber\\
\equiv  \int dt \mathcal{L} \equiv \int dt\left[\frac{1}{2}\mathcal{G}_{AB}\dot{q}^A\dot{q}^B-\mathcal{V}_0(q^1)^3V(q^2)\right],
\eea
where $\kappa = 8 \pi G$, $\mathcal{V}_0=\int_{\mathcal{M}^3}d^3x$, and $q^A=(q^1,q^2):=(a,\phi)$. Here we have assumed that the spatial section $\mathcal{M}^3$ of the universe is compact possesing the comoving volume $\mathcal{V}_0a^3$. Because of the restriction of the infinitely many degrees of freedom of superspace to only two, one arrives at a two-dimensional \textit{minisuperspace} defined by the metric 
\bea
\mathcal{G}_{AB}=\left(\begin{array}{cc}
-\frac{6\mathcal{V}_0}{\kappa}q^1 & 0\\
0	& \mathcal{V}_0(q^1)^3\\
\end{array}
\right)
\eea
in terms of which the Lagrangian density $\mathcal{L}=\mathcal{L}(q^A,\dot{q}^A)$ can be read of from~\eqref{action}. From $\mathcal{L}$ one obtains the canonical momenta $\pi_A=(\pi_1,\pi_2):=(\pi_a,\pi_{\phi})$
\bea
\lb{momenta}
\pi_1:= \frac{\partial \mathcal{L}}{\partial \dot{q}^1} &=-\frac{6\mathcal{V}_0}{\kappa}a\dot{a},\nonumber\\
\pi_2:= \frac{\partial \mathcal{L}}{\partial \dot{q}^2}& = \mathcal{V}_0a^3\dot{\phi}
\ .
\eea
The canonical Hamiltonian $\mathcal{H}$ is then given by
\bea
\lb{Ham}
\mathcal{H}(q^A,\pi_A)&:=\frac{1}{2} \mathcal{G}^{AB}\pi_A \pi_B +\mathcal{V}_0\left(q^1\right)^3V(q^2)
\ ,
\eea
where $\mathcal{G}^{AB}$ denotes the inverse metric. Explicitly, one obtains
\be
\lb{Ham2}
\mathcal{H}(a,\pi_a,\phi,\pi_{\phi}) = -\frac{\kappa}{12\mathcal{V}_0a}\pi_a^2+\frac{1}{2\mathcal{V}_0a^3}\pi_{\phi}^2+\mathcal{V}_0a^3V(\phi)
\ . 
\ee
The canonical Hamiltonian $\mathcal{H}$ is constrained to vanish, and expressing $\pi_a$ and $\pi_{\phi}$ in~\eqref{Ham2} in terms of $a$, $\dot{a}$, $\phi$, $\dot{\phi}$ via~\eqref{momenta} the Hamiltonian constraint becomes identical to the Friedmann equation
\bea
\lb{Friedmann}
H^2=\frac{\kappa}{3}\epsilon=\frac{\kappa}{3}\left(\frac{\dot{\phi}^2}{2} + V(\phi)\right),
\eea
where $H:=\frac{\dot{a}}{a}$ denotes the Hubble parameter. In addition, we have the equation of motion of the scalar field $\phi$, i.e. the Klein-Gordon equation
\be
\lb{KleinGordon2}
\ddot\phi+3H\dot\phi+\frac{\d V(\phi)}{\d\phi}=0 \ ,
\ee
which is automatically fulfilled if the energy density $\epsilon$ obeys the equation of energy conservation 
\be
\lb{continuity}
\dot{\epsilon}=-3H(\epsilon+p)\ .
\ee

\section{The classical big bang model}
\lb{klass}
In the following, we assume that the perfect fluid mimicked by the scalar field $\phi$ obeys the equation of state of a Chaplygin gas \cite{Chap_04}
\be
\lb{state}
p=p(\epsilon)=-\frac{A}{\epsilon}\ ,
\ee
where $A$ is a positive constant of dimension energy density squared. Inserting \eqref{state} into the continuity equation~\eqref{continuity}, one obtains $\epsilon$ in terms of the cosmic scale factor~$a$
\be
\lb{density}
\epsilon=\epsilon(a) = \sqrt{A+\frac{B}{a^6}} \ ,
\ee
where $B:=\left(\epsilon_0^2-A\right)a_0^6$, and $\epsilon_0$ resp. $a_0$ denote the energy density resp. the scale factor at the present epoch. From the Friedmann equation~\eqref{Friedmann} follows that $\epsilon_0$ is identical to the critical energy density $\epsilon_{\mathrm{crit}}:=\frac{3H_0^2}{\kappa}$ which is completely determined by Newton's gravitational constant $G$ and the Hubble constant $H_0$. Requiring that $\epsilon$ stays real and positive for all values of $a\geq 0$, we obtain $0 < A < \epsilon_0^2$. (The case $A\equiv0$ corresponds to a pure dust model, whereas the case $A\equiv\epsilon_0^2$ respectively $B\equiv 0$ corresponds to a pure cosmological constant.) Obviously, the energy density $\epsilon$ interpolates between a pure dust model with $\epsilon=\frac{\sqrt{B}}{a^3}$ and $p=0$ at the big bang, i.e. for $a\rightarrow0$, and a positive cosmological constant $\Lambda:=\kappa\sqrt{A}$ with equation of state $w_{\Lambda}:= \frac{p_{\Lambda}}{\epsilon_{\Lambda}}=-1$ in the limit $a\rightarrow \infty$. Introducing the dimensionsless parameter $\alpha$, $0 < \alpha < 1$, by $A=:\alpha \epsilon_0^2$, equation~\eqref{density} takes the simple form
\be
\lb{density_N}
\frac{\epsilon}{\epsilon_0}= \sqrt{\alpha+\left(1-\alpha\right)\left(\frac{a_0}{a}\right)^6} 
\ee
as a function of $\frac{a_0}{a}=1+z$, where $z$ is the redshift. In Fig.~\ref{figDENSITY} we show $\frac{\epsilon}{\epsilon_0}$ for $\alpha=0.897$. As a function of $z$ the equation of state $w(z):=\frac{p}{\epsilon}$, is given by
\be
\lb{w}
w(z)=-\frac{\alpha}{\alpha+(1-\alpha)(1+z)^6} \ . 
\ee
Using \eqref{density_N} one gets from the Friedmann equation~\eqref{Friedmann} for the expansion rate of the universe
\be
\lb{apunkt}
\frac{\dot{a}}{\dot{a}_0}=\frac{a}{a_0}\left[\alpha+\left(1-\alpha\right)\left(\frac{a_0}{a}\right)^6\right]^\frac{1}{4} \ , 
\ee
where $\dot{a}_0=a_0H_0$ denotes the expansion rate at the present epoch. In Fig.~\ref{da} we present $\frac{\dot{a}}{\dot{a}_0}$ for $\alpha=0.897$.

Combining the two basic equations~\eqref{Friedmann} and \eqref{continuity} implies quite generally the following equation 
\be
\lb{app_G}
\ddot{a}=-\frac{\kappa}{6}\left(\epsilon+3p\right)a \, 
\ee
for the acceleration of the universe which in the case of the Chaplygin gas reads 
\be
\lb{app}
\frac{\ddot{a}}{a_0H_0^2}=\left(\frac{a}{a_0}\right)\frac{\alpha-\frac{1-\alpha}{2}\left(\frac{a_0}{a}\right)^6}{\sqrt{\alpha+\left(1-\alpha\right)\left(\frac{a_0}{a}\right)^6}} \ .
\ee
One concludes from \eqref{app} that the universe expands faster and faster ($\ddot{a} > 0$) for redshifts obeying $z < (\frac{2\alpha}{1-\alpha})^{\frac{1}{6}}-1$. In order to obtain an accelerated expansion of the universe at the present epoch, one requires $\alpha > \frac{1}{3}$. Fig.~\ref{dda} shows $\frac{\ddot{a}}{a_0 H_0^2}$ as a function of $\frac{a}{a_0}$ for $\alpha=0.897$.

Integration of \eqref{apunkt} yields for small scale factors $a(t)$ $\left(a(t)<(\frac{B}{A})^{\frac{1}{6}}=(\frac{1-\alpha}{\alpha})^{\frac{1}{6}}a_0\right)$
\bea
\lb{st}
\sqrt{\frac{\kappa}{3}}B^{\frac{1}{4}}t&=&\int_0^{a(t)}\frac{\sqrt{\tilde{a}}\,d\tilde{a}}{\left(1+\frac{A}{B}\tilde{a}^6\right)^\frac{1}{4}}\nonumber\\
&=&\frac{2}{3}a(t)^{\frac{3}{2}}-\frac{A}{30B}a(t)^{\frac{15}{2}}+\mathcal{O}\left(a(t)^{\frac{27}{2}}\right),
\eea
from which one obtains the following expansion for the scale factor $a(t)$ as a function of cosmic time $t$ near to the big bang $(t\rightarrow0+)$
\be
\lb{01}
a(t)=a_0\left[\left(\frac{t}{t_1}\right)^{\frac{2}{3}}+\frac{1}{30}\frac{\alpha}{1-\alpha}\left(\frac{t}{t_1}\right)^{\frac{14}{3}}+\mathcal{O}\left(t^{\frac{25}{3}}\right)\right] \ , 
\ee
where $t_1:=\frac{2}{3}\left(1-\alpha\right)^{-\frac{1}{4}}t_H$ and $t_H:=\frac{1}{H_0}$ denotes the Hubble time.
As expected, the leading term in the expansion \eqref{01} agrees with the time-dependence of a pure dust model.

It is worthwhile to present an alternative relation for the scale factor valid at all times $0\leq t<\infty$. Let us rewrite \eqref{st} as follows
\be
\lb{alter}
\sqrt{\frac{\kappa}{3}}A^{\frac{1}{4}}t=\int_0^{a(t)}\frac{d\tilde{a}}{\tilde{a}\left(1+\frac{B}{A\tilde{a}^6}\right)^{\frac{1}{4}}}=\frac{2}{3}\int^{\infty}_{U(a)}\frac{ x^2 dx}{x^4-1}\ ,
\ee
where we have made the substitution $x:=\left(1+\frac{B}{A\tilde{a}^6}\right)^{\frac{1}{4}}$ with $U(a):=\left(1+\frac{1-\alpha}{\alpha}\left(\frac{a_0}{a}\right)^6\right)^{\frac{1}{4}}$.
We then obtain
\be
\lb{alter*}
\frac{t}{t_H}=\frac{1}{3\alpha^{\frac{1}{4}}}\left[\mathrm{arcoth}\,U(a) +\arctan\left(\frac{1}{U(a)}\right)\right].
\ee
This relation gives for the age of the universe
\be
\lb{age}
t_0=\frac{1}{3\alpha^{\frac{1}{4}}}\left[\frac{1}{2}\ln\left(\frac{1+\alpha^{\frac{1}{4}}}{1-\alpha^{\frac{1}{4}}}\right)+\arctan\left(\alpha^{\frac{1}{4}}\right)\right]t_H \ .
\ee
At large times, $t\rightarrow \infty$, one derives from \eqref{alter*} an exponential increase of the scale factor 
\be
\lb{exp}
a(t)=a_0\left(\frac{1-\alpha}{8\alpha}\right)^{\frac{1}{6}} \exp\left[\alpha^{\frac{1}{4}}\left(\frac{t}{t_H}\right)\right]+...
\ee
in complete agreement with the expected behavior in this limit due to a cosmological constant $\Lambda=\kappa\sqrt{A}$ yielding $\sqrt{\frac{\Lambda}{3}}=\frac{\alpha^{\frac{1}{4}}}{t_H}$. In Fig.~\ref{a} we display the evolution of the scale factor as a function of cosmic time for $\alpha=0.897$ belonging to an age of the universe $t_0=t_H$.

It remains to compute the cosmic evolution of the scalar field $\phi$ of the Chaplygin gas and the potential $V(\phi)$ associated with it. It turns out to be sufficient to know $\phi=\phi(a)$ as a function of the scale factor $a$. From \eqref{ep} follows $\dot{\phi}=\mp \sqrt{\epsilon+p}$, and using $\frac{d\phi}{da}=\frac{\dot{\phi}}{\dot{a}}$, the Friedmann equation~\eqref{Friedmann} and the equation of state~\eqref{state} for the Chaplygin gas combined with the explicit expression \eqref{density} for $\epsilon$, one obtains after one integration with $\phi(\infty)=0$
\be
\lb{phi*}
\phi(a)= \pm \frac{1}{\sqrt{3\kappa}}\mathrm{arcoth}\xi
\ee
with 
\be
\lb{xi}
\xi=\xi(a):= \sqrt{1+\frac{A}{B}a^6}=\sqrt{1+\frac{\alpha}{1-\alpha}\left(\frac{a}{a_0}\right)^6} \ .
\ee
From \eqref{phi*} and \eqref{xi} one derives for $a\rightarrow0$ the asymptotic
behavior
\bea
\lb{phias}
\phi(a)&=& \mp \sqrt{\frac{3}{\kappa}}\ln\left(\frac{a}{a_0}\right)\mp\frac{1}{2\sqrt{3\kappa}}\ln\left(\frac{\alpha}{4(1-\alpha)}\right)\nonumber\\
&+&\mathcal{O}\left(\left(\frac{a}{a_0}\right)^6\right) \ ,
\eea
and we conclude that $\phi(a)\rightarrow\pm\infty$ near to the big bang.
At large times, $a\rightarrow\infty$, the scalar field vanishes like
\be
\lb{phiasl}
\phi(a)= \pm \sqrt{\frac{1-\alpha}{3\kappa\alpha}}\left(\frac{a_0}{a}\right)^3+\mathcal{O}\left(\left(\frac{a_0}{a}\right)^9\right) \ .
\ee
In Fig.~\ref{phi} we present the two branches of $\phi(a)$ for $A=0.897\epsilon_0^2$.

To derive the corresponding scalar field potential $V$, we use 
\be
\lb{V}
V=\frac{1}{2}\left(\epsilon-p\right)=\frac{1}{2}\left(\epsilon+\frac{A}{\epsilon}\right)
\ee
and
\be
\lb{V*}
\epsilon(\phi)=\sqrt{A} \cosh\left(\sqrt{3\kappa}\phi\right)
\ee
which follows from \eqref{density},\eqref{phi*} and \eqref{xi}, and obtain
\be
\lb{VCH}
V(\phi)=\frac{\sqrt{A}}{2}\left[\cosh\left(\sqrt{3\kappa}\phi\right)+\frac{1}{\cosh\left(\sqrt{3\kappa}\phi\right)}\right]\ .
\ee

Note that the energy density $\epsilon(\phi)$, Eq.~\eqref{V*}, and the potential $V(\phi)$, Eq.~\eqref{VCH}, depend only on $\kappa$ and $A$ and no more on the integration constant $B$ reflecting the equation of state~\eqref{state} as it should.
The potential $V(\phi)$ is shown for $\phi>0$ in Fig.~\ref{potential} for $A=0.897\epsilon_0^2$.

\begin{figure}[t!]
\begin{center}
\includegraphics[width=7.5cm, angle=0]{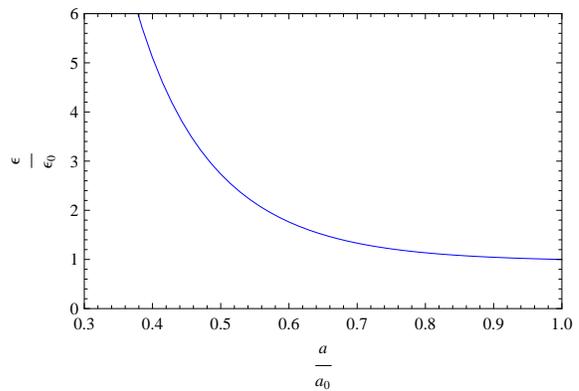}
\caption{The energy density $\epsilon$ of the scalar field (Chaplygin gas) as a function of the cosmic scale factor $a$ for $A=0.897\epsilon_0^2$.}
\lb{figDENSITY}
\end{center}
\end{figure}

\begin{figure}[t!]
\begin{center}
\includegraphics[width=7.5cm, angle=0]{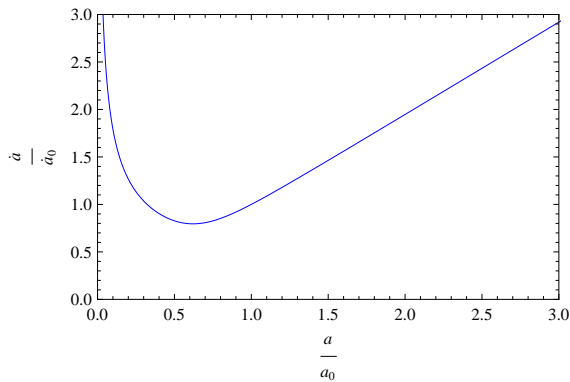}
\caption{The expansion rate $\dot{a}$ of the universe as a function of the cosmic scale factor $a$ for $A=0.897\epsilon_0^2$.}
\lb{da}
\end{center}
\end{figure} 

\begin{figure}[t!]
\begin{center}
\includegraphics[width=7.5cm, angle=0]{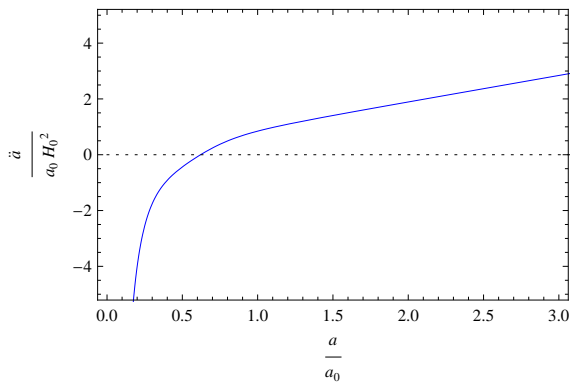}
\caption{Evolution of the cosmic acceleration $\ddot{a}$ as a function of the scale factor $a$ for $A=0.897\epsilon_0^2$.}
\lb{dda}
\end{center}
\end{figure}

\begin{figure}[t!]
\begin{center}
\includegraphics[width=7.5cm, angle=0]{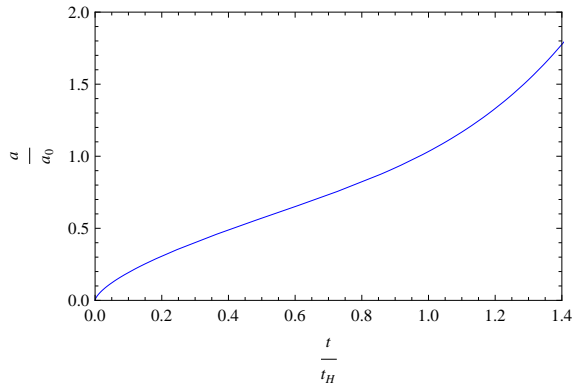}
\caption{Evolution of the scale factor $a(t)$ as a function of cosmic time $t$ for $A=0.897\epsilon_0^2$.}
\lb{a}
\end{center}
\end{figure} 

\begin{figure}[ht]
\begin{center}
\includegraphics[width=7.5cm, angle=0]{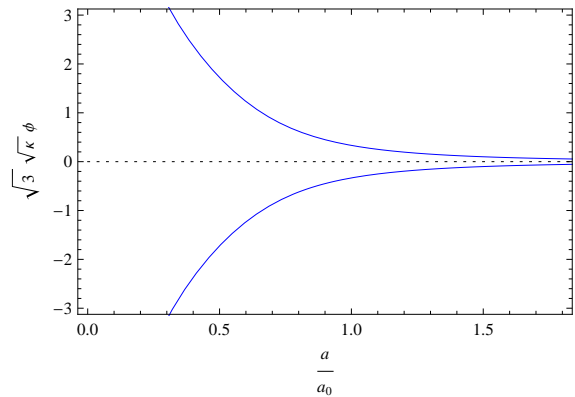}
\caption{The two branches of the scalar field $\phi$ as a function of $a$ for $A=0.897\epsilon_0^2$.}
\lb{phi}
\end{center}
\end{figure} 

\begin{figure}[ht]
\begin{center}
\includegraphics[width=7.5cm, angle=0]{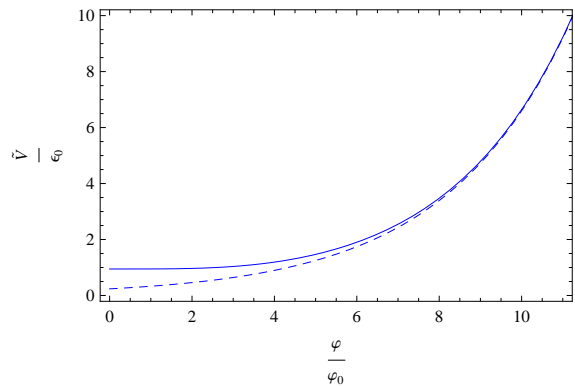}
\caption{The scalar field potential $\tilde{V}$ corresponding to the Chaplygin gas depicted over $\varphi$ for the branch $\varphi>0$ and $A=0.897\epsilon_0^2$. The dashed curve shows the approximation $\tilde{V}_{\mathrm{bb}}:= \frac{\sqrt{A}}{4}\exp\left(3\sqrt{2}\varphi\right)$ for $\varphi\rightarrow\infty$.}
\lb{potential}
\end{center}
\end{figure}


\section{The quantum big bang model}
\lb{sec:quant}
In this Section, we shall describe the quantization of the classical scenario discussed above. Quantization is carried out in the canonical approach in the Schr\"odinger picture. Implementing the Hamiltonian constraint via Dirac's quantization program of constrained systems, one arrives at the Wheeler-DeWitt equation \cite{DeWitt_67,Wheeler:64} depending on the configuration space variables
$(a,\phi)$ (respectively $(x,\varphi)$ to be defined below). From the solution obtained in this Section, we shall retrieve information concerning the above Chaplygin gas model with regard to its quantum behavior at the classical singularity and in the semiclassical limit. A detailed discussion of the influence of the boundary conditions on the wave function is presented in Sec.~\ref{BB}.

\subsection{Solution to the Wheeler-DeWitt equation in the vicinity of the big bang}\lb{sec:bb}
Now we analyze the quantum version of the classical model discussed in Sec. \ref{klass}. The wave function satisfies the Wheeler-DeWitt equation 
\be
\lb{WDW}
\left[\frac{\hbar^2}{2}\left(\frac{\partial^2}{\partial x^2}-\frac{\partial^2}{\partial\varphi^2}\right)+\frac{6}{\kappa}\mathcal{V}_0^2a_0^6\mathrm{e}^{6x}\tilde{V}(\varphi)\right]\Psi\left(x,\varphi\right)=0 \ ,
\ee
where we have introduced the dimensionless variables $x:=\ln{\frac{a}{a_0}}$ and $\varphi:= \sqrt{\frac{\kappa}{6}}\phi$ and used the Laplace-Beltrami factor ordering. The potential $\tilde{V}(\varphi)$ is given by $\tilde{V}(\varphi):=V\left(\sqrt{\frac{6}{\kappa}}\phi\right)$, see equation~\eqref{VCH}. 

In this Section, we are interested in the vicinity of the big bang, where one has $\left|\varphi\right|\rightarrow \infty$ and thus we can approximate the potential by an exponential function, see Fig.~\ref{potential}. In the following, we study the case $\varphi\geq 0$. Analogous results hold for the branch $\varphi \leq 0$. Thus we obtain near to the big bang the Wheeler-DeWitt equation 
:

\bea
\lb{WDW*}
\frac{\hbar^2}{2}\left(\frac{\partial^2}{\partial x^2}-\frac{\partial^2}{\partial\varphi^2}\right)\Psi_{\mathrm{bb}}\left(x,\varphi\right)\nonumber\\
+\frac{3\sqrt{A}}{2\kappa}\mathcal{V}_0^2a_0^6\exp\left[3\left(2 x+ \sqrt{2}\varphi\right)\right]\Psi_{\mathrm{bb}}\left(x,\varphi\right)=0 .
\eea
Here the index $\mathrm{bb}$ denotes the approximation to the wave function near to the big bang ($\mathrm{bb}$). We introduce new coordinates $u=u(x,\varphi), v=v(x,\varphi)$ (see also \cite{KAM_07}) in order to reduce Eq.~\eqref{WDW*} to 
\be
\lb{WDW**}
\hbar^2\left(\frac{\partial^2}{\partial u^2}-\frac{\partial^2}{\partial v^2}\right)\Psi_{\mathrm{bb}}\left(u,v\right)+\Psi_{\mathrm{bb}}\left(u,v\right)=0 \ ,
\ee
where
\be
\lb{u}
u\left(x,\varphi\right):=\frac{\lambda}{6}\left[\rho\mathrm{e}^{\sigma(x+\varphi)}+\sigma\mathrm{e}^{\rho(x-\varphi)}\right]\ ,
\ee
\be
\lb{v}
v\left(x,\varphi\right):=\frac{\lambda}{6}\left[\rho\mathrm{e}^{\sigma(x+\varphi)}-\sigma\mathrm{e}^{\rho(x-\varphi)}\right]
\ee
and
\bea
\lb{nc}
\lambda &:= &\frac{2A^{\frac{1}{4}}}{\sqrt{3\kappa}}\mathcal{V}_0a_0^3,\nonumber\\
\rho&:=& 3\left(1-\frac{1}{\sqrt{2}}\right),\\
\sigma&:=& 3\left(1+\frac{1}{\sqrt{2}}\right).\nonumber
\eea
The solution to \eqref{WDW**} can be found from the \textit{ansatz}
\be\lb{steineranstaz}
\Psi_\mathrm{bb}\left(u,v\right)=f(\gamma_1 u+ \gamma_2 v) \ ,
\ee
where $\gamma_1$ and $\gamma_2$ do not depend on $(u,v)$. Inserting \eqref{steineranstaz} into the Wheeler-DeWitt equation~\eqref{WDW**} leads to  
\be
\hbar^2\left(\gamma_1^2 - \gamma_2^2\right)f''+f=0
\ , 
\ee
where the prime denotes a derivative with respect to the whole argument of $f$. Setting $\hbar^2\left(\gamma_1^2 - \gamma_2^2\right)=1$, we get with $\gamma_2:=\pm\frac{k}{\hbar}$, $k\geq0$, $\gamma_1=\frac{\omega}{\hbar}$ and $\omega=\omega(k):=+\sqrt{k^2+1}$ the simple differential equation
\be
\lb{k_0}
f''+f=0
\ , 
\ee
which has the two solutions ($\mathcal{C}_{1/2}^\pm$ are arbitrary complex coefficients)
\bea
\lb{Psi0}
\Psi_{\mathrm{bb}}^\pm(u,v) = \mathcal{C}_1^\pm\sin\,\left(\frac{S_\pm}{\hbar}\right)+\mathcal{C}_2^\pm\cos\,\left(\frac{S_\pm}{\hbar}\right).
\eea
Here the functions 
\be
\lb{actionbb}
S_\pm=S_\pm(k;u,v):=\omega (k) u\pm k v
\ee
denote the classical actions satisfying the Hamilton-Jacobi equation corresponding to the Wheeler-deWitt equation~\eqref{WDW**}. Although the solutions~\eqref{Psi0} solve the Wheeler-DeWitt equation for every complex value of $k$ (with $\omega(k)$ being then complex, too), we have chosen the parameter $k$ to be real and positive, since the physical boundary conditions to be discussed in the next Sections necessarily lead to the interpretation of $k$ as a real wave number. Furthermore, we note that the two other admissible classical actions obtained by $S_\pm\rightarrow-S_\pm$ do not lead to new linearly independent solutions of \eqref{WDW**}. Thus we conclude that the complete solution of the Wheeler-DeWitt equation~\eqref{WDW**} is given by $\Psi_{\mathrm{bb}}^++\Psi_{\mathrm{bb}}^-$. In Sec~\ref{BB} we shall show that the coefficients $\mathcal{C}_{1/2}^\pm$ are determined by the physically correct boundary conditions . 


Here we want to point out that the exponential approximation to the exact Chaplygin gas potential~\eqref{VCH} does not interpolate between matter and dark energy, contrary to the Chaplygin gas. In fact, it mimics a pure dust component with $p=0$ respectively $w=0$ (see Sec.~\ref{dust}). This means, due to the fact that the pressure of the Chaplygin gas behaves like $p\sim -\frac{A}{\sqrt{B}}a^3+\ldots$ near to the big bang, that the exponential potential $\tilde{V}_{\mathrm{bb}}$ describes the Chaplygin gas precisely at the big bang, $a\equiv 0$. Including higher order terms in the expansion of the Chaplygin gas potential in the vicinity of the big bang would be interesting, but would lead to a more complicated Wheeler-DeWitt equation, whose exact solution has not yet been found.
   
\subsection{Solution to the Wheeler-DeWitt equation for large values of the scale factor}
\lb{sec:largevalues}

In the case of large values of the scale factor, $a\rightarrow\infty$, respectively $x\rightarrow\infty$, the classical field $\varphi$ approaches zero, see Eq.~\eqref{phiasl}, and therefore we can approximate the scalar potential $\tilde{V}(\varphi)$ in the Wheeler-DeWitt equation \eqref{WDW} by its leading term $\sqrt{A}$, since $\tilde{V}(\varphi)=\sqrt{A}\left(1+\mathcal{O}(\varphi^4)\right)$ for $\varphi\rightarrow 0 $. Thus we study in the following the Wheeler-DeWitt equation 
\bea
\label{WDW***}
\hbar^2\left(\frac{\partial^2}{\partial x^2}-\frac{\partial^2}{\partial\varphi^2}\right)\Psi_\infty\left(x,\varphi\right)\nonumber\\
+9\lambda^2 \mathrm{e}^{6 x}\Psi_\infty\left(x,\varphi\right)=0
\ , 
\eea
where the index $\infty$ denotes the case $a\rightarrow\infty$. 
To solve this equation, we make the \textit{separation ansatz}
\be
\Psi_\infty\left(x,\varphi\right)=C_\infty(x)\Phi_\infty(\varphi) 
\ee
and arrive at 
\be
\lb{eq}
\hbar^2\frac{C_\infty''(x)}{C_\infty(x)}+9\lambda^2\mathrm{e}^{6x}=\hbar^2\frac{\Phi_\infty''(\varphi)}{\Phi_\infty(\varphi)}\equiv-K^2 
\ee
respectively
\be
\Phi_\infty''(\varphi)+\left(\frac{K}{\hbar}\right)^2\Phi_\infty(\varphi)=0. 
\ee
Thus the solution for the matter-dependent part of the wave function $\Psi_\infty$ is given by
\bea
\lb{agmatter}
\Phi_\infty(\varphi) = \mathcal{A}_1 \mathrm{e}^{i\frac{K\varphi}{\hbar}}
+\mathcal{A}_2 \mathrm{e}^{-i\frac{K\varphi}{\hbar}}.
\eea
In order to solve the gravitational part of Eq.~\eqref{eq}
\be
\label{f''}
C_\infty''(x)+\left[9\left(\frac{\lambda}{\hbar}\right)^2 \mathrm{e}^{6x}+\left(\frac{K}{\hbar}\right)^2\right]C_\infty(x)=0,
\ee
we make the \textit{ansatz}
\be
\lb{an}
C_{\infty}(x)\equiv F(z), 
\ee
where $z:=\beta  \mathrm{e}^{3x}$.
Inserting Eq.~\eqref{an} into \eqref{f''}, one gets the Bessel differential equation \cite{MAGNUS_66}
\be
\lb{z}
z^2 F''(z)+ z F'(z)+\left(z^2-\nu^2\right)F(z)=0,
\ee
if $\beta$ is chosen to be equal to $\frac{\lambda}{\hbar}$ (see Eq.~\eqref{nc}) and $\nu$ is given by $i\frac{K}{3\hbar}$.
Thus we get the following solution for the gravitational part:
\bea
\lb{aggrav}
C_{\infty}(x)=\mathcal{K}_1J_{i\frac{K}{3\hbar}}\left(\frac{\lambda}{\hbar} \,\mathrm{e}^{3x}\right)
+\mathcal{K}_2 Y_{i\frac{K}{3\hbar}}\left(\frac{\lambda}{\hbar}\, \mathrm{e}^{3x}\right),
\eea
where $J_{\nu}(z)$ is the Bessel function and $Y_{\nu}(z)$ the Neumann function. We can rewrite the above wave function in terms of the Hankel functions $H_\nu^{(i)}(z)$ 
\bea
\lb{f}
C_{\infty}(x)=\mathcal{D}_1H^{(1)}_{i\frac{K}{3\hbar}}\left(\frac{\lambda}{\hbar} \,\mathrm{e}^{3x}\right)
+\mathcal{D}_2 H^{(2)}_{i\frac{K}{3\hbar}}\left(\frac{\lambda}{\hbar}\,\mathrm{e}^{3x}\right).
\eea
The general solution to the Wheeler-DeWitt equation~\eqref{WDW***} is then given by:
\be
\lb{Psiunend}
\begin{split}
\Psi_\infty(x,\varphi) &=\left[\mathcal{A}_1 \mathrm{e}^{i\frac{K\varphi}{\hbar}}+\mathcal{A}_2 \mathrm{e}^{-i\frac{K\varphi}{\hbar}}\right] \\
& \times  \left[\mathcal{D}_1H^{(1)}_{i\frac{K}{3\hbar}}\left(\frac{\lambda}{\hbar} \,\mathrm{e}^{3x}\right)
+\mathcal{D}_2 H^{(2)}_{i\frac{K}{3\hbar}}\left(\frac{\lambda}{\hbar}\,\mathrm{e}^{3x}\right)\right], 
\end{split}
\ee
with the arbitrary constants $\mathcal{A}_i\,,\mathcal{D}_{i}$ and $K$ which will be determined in Sec.~\ref{semisteiner}.
\begin{figure}[ht]
\begin{center}
\includegraphics[width=7.5cm, angle=0]{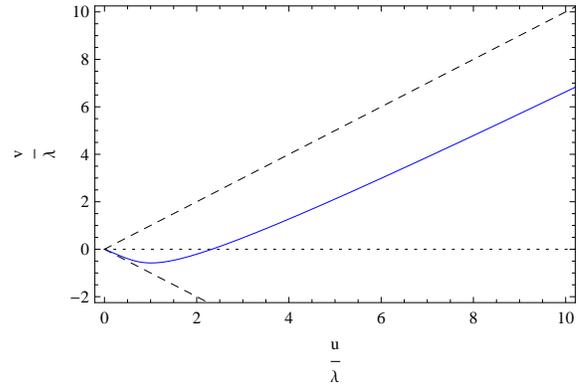}
\caption{The classical trajectory $\varphi_{\mathrm{cl}}$ (positive branch) as a function of the coordinates $u$ and $v$. The dashed lines $v=\pm u$ indicate the boundaries of the physically allowed sector.}
\lb{Fig:UVPaper}
\end{center}
\end{figure} 

\section{Avoiding the big bang singularity}
\lb{BB}
In Sec.~\ref{sec:bb} we have derived the general solution~\eqref{Psi0} to the Wheeler-DeWitt equation~\eqref{WDW*} respectively \eqref{WDW**} being an approximation valid near to the classical big bang to the exact Wheeler-DeWitt equation~\eqref{WDW}. This solution still depends on the four coefficients $\mathcal{C}_{1/2}^\pm$ and the constant $k$. It is the purpose of this Section to show that a physically sensible and mathematically consistent choice of the boundary conditions close to the big bang and in the far future exists, that avoids the classical big bang singularity and is strongly peaked exactly at the classical field solution in the semiclassical limit.

The solution~\eqref{Psi0} has been found in terms of the auxiliary variables $u$ and $v$ defined in Eqs.~\eqref{u} and \eqref{v}. In order to obtain their physically allowed range, one must go back to minisuperspace with the coordinates $(a,\varphi)\in\mathds{R}_>\times\mathds{R}_>$. (Remember, that we restrict the discussion to the positive branch of the scalar field $\varphi$). It then follows from \eqref{u} and \eqref{v} that the physically allowed domain of $u$ and $v$ is given by the sector $\mathcal{S}:=\{(u,v)\in\mathds{R}_>\times\mathds{R}, |v|\leq u \}$ whose boundary is indicated by the dashed lines in Fig.~\ref{Fig:UVPaper}. We also show in Fig.~\ref{Fig:UVPaper} the classical trajectory $\varphi_{\mathrm{cl}}:=\varphi_{\mathrm{cl}}(a(u,v))=\sqrt{\frac{\kappa}{6}}\phi(a(u,v))$ given in Eq.~\eqref{phi*} (with the upper sign) and shown in Fig.~\ref{phi} as a function of the scale factor $a$.

As is well-known, there exists no genuine time variable in the geometrodynamical approach. One can use, however, either the cosmic scale factor $a$ or the scalar field $\varphi$ as an "effective time" variable (emergent time). In the following, we consider the scale factor $a$ (respectively $x:=\ln\,(\frac{a}{a_0})$) to play the role of a "clock" for the quantum evolution. The deep quantum regime corresponding to the classical big bang is then defined as the hypersurface $a=0$, and thus the quantum evolution requires the definition of an initial condition imposed on the quantum wave function on this hypersurface ($a=0$). Here, we choose \textit{DeWitt's initial condition} on the hypersurface $a=0$ in the minisuperspace $(a,\varphi)\in\mathds{R}_>\times\mathds{R}_>$
\bea
\lb{cond1}
\lim_{a\rightarrow0}\Psi(a,\varphi)=0,
\eea
which should hold for any given value of the scalar field $\varphi\in\mathds{R}_>$. Note that we do not require that $\varphi$ has to be evaluated on the classical trajectory $\varphi_{\mathrm{cl}}(a)$. However, the condition~\eqref{cond1} does not completely determine the wave function, but rather has to be supplemented by an asymptotic boundary condition controlling the behavior of $\Psi$ in the far future, $a\rightarrow\infty$, as discussed below. 

We are now in the position to give the initial condition on the wave function $\Psi(u,v)$, Eq.~\eqref{Psi0}, considered as a function of $u$ and $v$, by translating the initial condition \eqref{cond1} into the $(u,v)$-plane. From \eqref{u} and \eqref{v} one obtains $u=\mathcal{O}\left((\frac{a}{a_0})^\rho\right)\rightarrow 0$ and $v=-u+\mathcal{O}\left((\frac{a}{a_0})^\sigma\right)\rightarrow 0$ for $a\rightarrow 0$, $\varphi$ fixed, and thus the initial condition becomes 
\be
\lb{cond1*}
\lim_{u\rightarrow0^+}\Psi(u,-u)=0.
\ee
Using the solutions $\Psi_{\mathrm{bb}}^\pm(u,v)$, Eq.~\eqref{Psi0}, the initial condition \eqref{cond1*} gives $\mathcal{C}_2^+=\mathcal{C}_2^-=0$, and we obtain
\bea
\lb{Psi0*}
\Psi_{\mathrm{bb}}(u,v)&=&\mathcal{C}_1^+ \sin\left(\frac{S_+}{\hbar}\right)+\mathcal{C}_1^-  \sin\left(\frac{S_-}{\hbar}\right)
\eea
as the solution of the Wheeler-DeWitt equation~\eqref{WDW**} satisfying DeWitt's criterium~\eqref{cond1} for avoiding the classical big bang singularity.

Let us remark that the wave function \eqref{Psi0*} satisfies the Wheeler-DeWitt equation near to the classical big bang together with the initial condition \eqref{cond1} for \textit{any} complex value of $k$. This is important since the value of $k$ respectively its admissible values (discrete or possibly continuous) cannot be determined from a solution whose validity is restricted to small values of the scale factor $a$. To completely fix the spectrum of $k$ (and thus of $\omega$) requires the knowledge of the wave function in the whole minisuperspace which in turn can only be determined (at least in principle) if the allowed space of functions equipped with an appropriate scalar product is given. While we shall present in Section \ref{semisteiner} the quantum wave function satisfying the approximate Wheeler-DeWitt equation valid in the classical regime, i.e. for $a\rightarrow\infty$, and thus for late times, the complete solution for the Chaplygin gas valid in the whole minisuperspace is not known. 

As already mentioned, the initial condition~\eqref{cond1} has to be supplemented by a second boundary condition. Here, we propose the following \textit{semiclassical boundary condition} which selects from a wave packet based on the wave function \eqref{Psi0*} one which in the limit of $\hbar\rightarrow 0$ is strongly peaked at the classical field configuration $\varphi_{\mathrm{cl}}^+$ corresponding to the positive branch of the scalar field. (Remember that in the Wheeler-DeWitt equation~\eqref{WDW*} we have used the approximation $\tilde{V}_{\mathrm{bb}}$ to the scalar potential valid for $\varphi\geq0$). Since the Wheeler-DeWitt equation is linear, we can form from \eqref{Psi0*} the most general wave packet using the superposition principle and obtain
\be
\begin{split}
\lb{superposi}
\Psi_{\mathrm{bb}}(u,v)&=\int_{0}^{\infty}\mathcal{C}_1^+(k) \sin\left(\frac{S_+}{\hbar}\right)\frac{dk}{\sqrt{2\pi\hbar}}\\&+\int_{0}^{\infty}\mathcal{C}_1^-(k) \sin\left(\frac{S_-}{\hbar}\right)\frac{dk}{\sqrt{2\pi\hbar}}\ .
\end{split}
\ee
(Here we have assumed that the spectrum of $k$ is continuous and given by $k\geq 0$). Note that Eq.~\eqref{superposi} displays explicitly the dependence on $\hbar$, since the classical actions $S_\pm$ are independent of $\hbar$. The crucial point is that the wave packet~\eqref{superposi} still satisfies the Wheeler-DeWitt equation~\eqref{WDW**} with the initial condition \eqref{cond1*} for any given complex "wave number amplitude" $\mathcal{C}_{1/2}^+(k)$. 

Rewriting~\eqref{superposi} as follows
\be
\begin{split}
\lb{superposi*}
\Psi_{\mathrm{bb}}(u,v)&=\int_{0}^{\infty}\mathcal{C}_1^+(k) \left[\mathrm{e}^{i\frac{S_+}{\hbar}}-\mathrm{e}^{-i\frac{S_+}{\hbar}}\right]\frac{dk}{2i\sqrt{2\pi\hbar}}\\&+\int_{0}^{\infty}\mathcal{C}_1^-(k) \left[\mathrm{e}^{i\frac{S_-}{\hbar}}-\mathrm{e}^{-i\frac{S_-}{\hbar}}\right]\frac{dk}{2i\sqrt{2\pi\hbar}}\ ,
\end{split}
\ee
we observe that \eqref{superposi*} has the standard form of a WKB-wave function which often is used as an \textit{ansatz} to obtain an approximate solution for quantum wave functions in the semiclassical limit when $\hbar$ goes to zero. Thus the wave function \eqref{superposi*} has a simple physical interpretation in terms of the classical actions $S_{\pm}$ which justifies our assumption of $k$ being real. It is important, however, to observe that \eqref{superposi*} has been derived from our exact solution \eqref{superposi} to the Wheeler-DeWitt equation~\eqref{WDW**} satisfying the crucial initial condition \eqref{cond1*} without assuming the WKB-approximation. We would like to mention that a related wave function has been introduced in \cite{KAM_07} in the discussion of the anti-Chaplygin gas, in that case, however, as an WKB-\textit{ansatz} which does not satisfy the initial condition \eqref{cond1*}.

In order to investigate the limit $\hbar\rightarrow 0$ of the wave packet~\eqref{superposi*}, one uses the method of stationary phase (see the detailed discussions in Sections \ref{semisteiner} and \ref{dust}) and finds that (under suitable conditions on the functions $\mathcal{C}_1^\pm(k)$) the first integral is in the semiclassical limit strongly peaked at the positive branch of the classical field and the second one at the negative branch. Employing now the \textit{asymptotic boundary condition} that the wave packet should reproduce only the classical result corresponding to the positive branch, we conclude $\mathcal{C}_1^-(k)\equiv 0$, and thus the \textit{exact} wave packet solution to the Wheeler-DeWitt equation~\eqref{WDW**} satisfying both boundary conditions takes the simple form
\be
\lb{superposi**}
\Psi_{\mathrm{bb}}(u,v)=\int_{0}^{\infty} \mathcal{C}(k) \left[\mathrm{e}^{i\frac{S_+}{\hbar}}-\mathrm{e}^{-i\frac{S_+}{\hbar}}\right]\frac{dk}{\sqrt{2\pi\hbar}}\ ,
\ee
where we have redefined the ampliude $\mathcal{C}(k)$.

The details of the calculation using the method of stationary phase will not be discussed at this point, since the derivation is identical to the one to be presented in Sec.~\ref{dust}, where we shall study a pure dust model instead of the Chaplygin gas. Since the Wheeler-DeWitt equation can be exactly solved for all values of $x$ in the case that the matter content of the universe consists only of a dust component, we are then able to discuss the exact wave function in the whole kinematically allowed range.

In conclusion, we have demonstrated that the Wheeler-DeWitt equation possesses near to the classical singularity a unique solution corresponding to the quantum wave function \eqref{superposi**} which avoids the big bang singularity in the sense of DeWitt's criterium \eqref{cond1} respectively \eqref{cond1*} and is strongly peaked at the classical field $\varphi_{\mathrm{cl}}^+(x)$ in the semiclassical limit.

\section{Quantum Cosmology of the Chaplygin Gas in the semiclassical limit}\lb{semisteiner}
In equation~\eqref{Psiunend} we have already given the general solution $\Psi_{\infty}(x,\varphi)$ to the Wheeler-DeWitt equation for the Chaplygin gas in the small-field approximation. To obtain from this expression the leading term in the semiclassical limit, we have to investigate the behavior of the Hankel functions $H^{(1,2)}_{i\frac{K}{3\hbar}}\left(\frac{\lambda}{\hbar}\,\mathrm{e}^{3x}\right)$ for $\hbar\rightarrow 0$ and $x,K$ fixed. This limit corrsponds to the Debye type asymptotics of $H^{(1,2)}_{i\mu}\left(z\right)$, where both $\mu:=\frac{K}{3\hbar}$ and $z:=\frac{\lambda}{\hbar}\,\mathrm{e}^{3x}$ are large and real positive numbers such that $\frac{\mu}{z}=\frac{K}{3\lambda}\mathrm{e}^{-3x}$ is fixed as $\mu,z\rightarrow \infty$. Using the well-known asymptotic formula in this limit (see \cite{MAGNUS_66}), we obtain for $\hbar\rightarrow 0$
\be\lb{PsiStein}
\begin{split}
\Psi_{\infty}(x,\varphi)&=\frac{1}{\left(K^2+m^2(x)\right)^{\frac{1}{4}}}\\
&\times \left[\mathcal{B}_1\mathrm{e}^{\frac{i}{\hbar}S_\infty^+}+\mathcal{B}_2\mathrm{e}^{-\frac{i}{\hbar}S_\infty^+}
\right.\\
&\left.+\mathcal{B}_3\mathrm{e}^{\frac{i}{\hbar}S_\infty^-}+\mathcal{B}_4\mathrm{e}^{-\frac{i}{\hbar}S_\infty^-}\right]
\left(1+\mathcal{O}(\hbar)\right).
\end{split}
\ee
Here the functions 
\be\lb{SStein}
\begin{split}
S^{\pm}_{\infty}=S^{\pm}_{\infty}(K;x,\varphi):=&\frac{1}{3}\sqrt{K^2+m^2(x)}\\-&\frac{K}{3}\mathrm{arsinh}\left(\frac{K}{m(x)}\right)\pm K\varphi
\end{split}
\ee
denote the two classical actions satisfying the Hamilton-Jacobi equation corresponding to the Wheeler-DeWitt equation~\eqref{WDW***}. Furthermore, we have defined 
\be
m(x):=3\lambda\mathrm{e}^{3x}
\ee 
and have introduced the new ($K$-dependent) coefficients
$\mathcal{B}_{1,2}:=\frac{\sqrt{6\hbar}}{\pi} \mathrm{e}^{\pm\frac{\pi K}{6\hbar}\mp i\frac{\pi}{4}}\mathcal{A}_{1,2}\mathcal{D}_{1,2}$ and $\mathcal{B}_{3,4}:=\frac{\sqrt{6\hbar}}{\pi} \mathrm{e}^{\pm\frac{\pi K}{6\hbar}\mp i\frac{\pi}{4}}\mathcal{A}_{2,1}\mathcal{D}_{1,2}$.
Since the Wheeler-deWitt equation~\eqref{WDW***} is linear, the leading term of the most general wave packet is given in the semiclassical limit by
\be\lb{PsiStein*}
\begin{split}
\Psi_{\infty}(x,\varphi)&=\int_{-\infty}^{\infty}\frac{dK}{\sqrt{2\pi\hbar}\left(K^2+m^2(x)\right)^{\frac{1}{4}}}\\&\times\left[\mathcal{B}_1(K)\mathrm{e}^{\frac{i}{\hbar}S_\infty^+}+\mathcal{B}_2(K)\mathrm{e}^{-\frac{i}{\hbar}S_\infty^+}
\right.\\&\left.+\mathcal{B}_3(K)\mathrm{e}^{\frac{i}{\hbar}S_\infty^-}+\mathcal{B}_4(K)\mathrm{e}^{-\frac{i}{\hbar}S_\infty^-}\right]\left(1+\mathcal{O}(\hbar)\right),
\end{split}
\ee
where $\mathcal{B}_i(K)$ now denote general complex "wave amplitudes". We shall now show, using the method of stationary phase, that the wave packet \eqref{PsiStein*} is for $\hbar\rightarrow0$ strongly peaked at the classical field trajectory $\varphi_{\mathrm{cl}}(x)$ if the amplitudes $\mathcal{B}_i(K)$ are strongly peaked at the "classical wave number" $K_{\mathrm{cl}}$ to be determined below.

In a first step let us assume that the amplitudes $\mathcal{B}_i(K)$ are smooth functions with compact support and that the actions $S^{\pm}_{\infty}$ possess a finite number of isolated non-degenerate stationary points $K^\pm_\gamma=K^\pm_\gamma(x,\varphi)\in \mathrm{supp}\,\mathcal{B}_i(K)$,
\be\lb{stationaryphase}
\begin{split}
\frac{\partial S^{\pm}_{\infty}}{\partial K}\left(K^\pm_\gamma(x,\varphi) ; x ,\varphi\right)&=0 \ ,\\
\frac{\partial^2 S^{\pm}_{\infty}}{\partial K^2}\left(K^\pm_\gamma(x,\varphi) ; x ,\varphi\right)&\not=0 \ .
\end{split}
\ee
We then obtain for the first term in the integral \eqref{PsiStein*}, denoted by $\Psi^{(1)}_\infty(x,\varphi)$, in the semiclassical limit, $\hbar\rightarrow 0$ \cite{Grigis:94}
\be
\lb{Psi1}
\begin{split}
\Psi^{(1)}_\infty(x,\varphi)&=\sum_{K_\gamma^+}\mathcal{B}_1\left(K^+_\gamma(x,\varphi)\right)\\&\times\frac{\left|\frac{\partial^2 S^{+}_{\infty}}{\partial K^2}\left(K^+_\gamma(x,\varphi) ; x ,\varphi\right)\right|^{-\frac{1}{2}}}{\left[\left(K_\gamma^+(x,\varphi)\right)^2+m^2(x)\right]^{\frac{1}{4}}}\\
&\times \exp\left\{\frac{i}{\hbar}S_\infty^+\left(K^+_\gamma(x,\varphi) ; x ,\varphi\right)\right.\\&\left.+\frac{i\pi}{4}\mathrm{sgn}\left[\frac{\partial^2 S^{+}_{\infty}}{\partial K^2}\left(K^+_\gamma(x ,\varphi) ; x ,\varphi\right)\right]\right\}\\&\times\left(1+\mathcal{O}(\hbar)\right).
\end{split}
\ee
For the remaining three terms of \eqref{PsiStein*} we obtain similar expressions with the obvious substitutions made in \eqref{Psi1}.

With the classical actions $S^{\pm}_{\infty}$ given in \eqref{SStein}, we obtain 
\be
\lb{S/K}
\frac{\partial S^{\pm}_{\infty}}{\partial K}=-\frac{1}{3}\mathrm{arsinh} \left( \frac{K}{m}\right)\pm \varphi
\ee
and
\be
\lb{S2/K2}
\frac{\partial^2 S^{\pm}_{\infty}}{\partial K^2}=-\frac{1}{3}\frac{1}{\sqrt{K^2+m^2}}
\ee
and thus the only stationary points are
\be
\lb{stationary points}
K_\gamma^\pm(x,\varphi)=\pm m (x) \sinh\left(3\varphi\right)
\ee
leading to
\be
\lb{S2/K2*}
\frac{\partial^2 S^{\pm}_{\infty}}{\partial K^2}\left(K_\gamma(x,\varphi) ; x ,\varphi\right)=- \frac{1}{3m (x) \cosh\left(3\varphi\right)}.
\ee

In the last step let us assume that the functions $\mathcal{B}_i(K)$ are strongly peaked at the classical wave number $K=K_{\mathrm{cl}}$ which in turn implies that the amplitudes $\mathcal{B}_i\left(K_\gamma^\pm(x,\varphi)\right)$ are strongly peaked at $K_\gamma^\pm (x,\varphi)=K_{\mathrm{cl}}$. From this condition and \eqref{stationary points} we conclude that e.g. the wave packet $\Psi^{(1)}_\infty(x,\varphi)$ derived in equation~\eqref{Psi1} is strongly peaked at those $(x,\varphi)$-values which are given by
\be
\lb{phicl}
\varphi^+_{\mathrm{scl}}(x):=\frac{1}{3}\mathrm{arsinh} \left( \frac{K_{\mathrm{cl}}}{m(x)}\right)=\frac{1}{3}\mathrm{arsinh} \left( \frac{K_{\mathrm{cl}}}{3\lambda}\mathrm{e}^{-3x}\right).
\ee
Since $\Psi^{(1)}_\infty(x,\varphi)$ is a solution to the Wheeler-DeWitt equation for the Chaplygin gas in the small field approximation respectively in the limit $x=\ln\,(\frac{a}{a_0})\gg 1$, we infer from \eqref{phicl} for the semiclassical field $\varphi^+_{\mathrm{scl}}(x)$ the asymptotic behavior 
\be
\lb{phiclasym}
\varphi^+_{\mathrm{scl}}(x)=  \frac{K_{\mathrm{cl}}}{9\lambda}\left(\frac{a_0}{a}\right)^3+\mathcal{O}\left(\left(\frac{a_0}{a}\right)^9\right), \quad a\gg a_0, 
\ee
in complete agreement with the exact behavior \eqref{phiasl} in this limit (for the positive branch of the scalar field) if the classical wave number $K_{\mathrm{cl}}$ takes the value
\be
\lb{K_cl}
K_{\mathrm{cl}}:=\sqrt{\frac{6B}{\kappa\sqrt{A}}}\mathcal{V}_0 \ .
\ee
Considering in a similar way the contributions to the wave packet $\Psi_\infty(x,\varphi)$ involving the action $S^-_\infty$, we conclude - using the same value \eqref{K_cl} for $K_{\mathrm{cl}}$ - that they are strongly peaked at $K_\gamma^-(x,\varphi)=K_{\mathrm{cl}}$ and thus at those ($x,\varphi$)-values which are given by 
\bea
\lb{phicl-}
\varphi^-_{\mathrm{scl}}(x)&:=&-\frac{1}{3}\mathrm{arsinh} \left( \frac{K_{\mathrm{cl}}}{3\lambda}\mathrm{e}^{-3x}\right)\\\nonumber
&=&-\frac{K_{\mathrm{cl}}}{9\lambda}\left(\frac{a_0}{a}\right)^3+\mathcal{O}\left(\left(\frac{a_0}{a}\right)^9\right), \quad a\gg a_0,
\eea
again in complete agreement with the exact asymptotic behavior \eqref{phiasl} for the negative branch of the scalar field.

It is interesting to compare the semiclassical results \eqref{phicl} and \eqref{phicl-} with the exact formula \eqref{phi*}, valid at all $a\ge 0$, which can be rewritten in terms of $K_{\mathrm{cl}}$ and reads (for the positive (+) respectively the negative (-) branch)
\be
\lb{phiclpm}
\varphi^\pm_{\mathrm{cl}}(x)=\pm\frac{1}{3\sqrt{2}}\mathrm{arsinh} \left( \sqrt{2}\frac{K_{\mathrm{cl}}}{3\lambda}\mathrm{e}^{-3x}\right).
\ee
While the leading terms in $\varphi^\pm_{\mathrm{scl}}$ and $\varphi^\pm_{\mathrm{cl}} $ agree, it is seen from \eqref{phiclpm} that the higher order terms in $(\frac{a_0}{a})$ are different.

We have thus demonstrated that there exists a quantum wave packet derived from the Wheeler-DeWitt equation which in the semiclassical limit is strongly peaked exactly at the classical field configurations.

\section{Quantum Cosmology of a pure dust model}
\lb{dust}

In the previous Sections, we have investigated quantum cosmology for a universe whose matter content is assumed to consist of an ideal fluid satisfying the equation of state of a Chaplygin gas. The latter is described by a homogeneous  scalar field whereby the dynamics is completely fixed by the scalar potential \eqref{VCH}. The Chaplygin gas interpolates between pure dust at the big bang and dark energy described by a positive cosmological constant at late cosmic times.

In this Section, we study a pure dust ($\mathrm{d}$) model satisfying for all scale factors $a\geq 0$ the equation of state $p_{\mathrm{d}}\equiv 0$, i.e. a pressureless fluid. We know that the energy content of our Universe is dominated during a long period of its evolution by dust consisting of baryonic and dark matter. The equation of energy conservation~\eqref{continuity} determines completely the $a$- respectively $x$-dependence of the energy density $\epsilon_{\mathrm{d}}$ for dust which reads
\be\lb{energydust}
\epsilon_{\mathrm{d}}=\epsilon_0\left(\frac{a_0}{a}\right)^3=\epsilon_0\mathrm{e}^{-3x}\ . 
\ee
Here $\epsilon_0$ is given, as in the previous Sections, in terms of the Hubble constant $H_0$ since the Friedmann equations~\eqref{Friedmann} leads to $\epsilon_0=\epsilon_{\mathrm{crit}}:=\frac{3H_0^2}{\kappa}$. Furthermore, it follows $V_{\mathrm{d}}=\frac{\epsilon_{\mathrm{d}}}{2}$ from the two equations in \eqref{ep} which then yields for the scalar potential
\be\lb{potentialdust}
\tilde{V}_{\mathrm{d}}\left(\varphi_{\mathrm{d}}(x)\right)=\frac{\epsilon_0}{2}\mathrm{e}^{-3x}\ , 
\ee
where we have introduced again the dimensionless field $\varphi_{\mathrm{d}}:=\sqrt{\frac{\kappa}{6}}\phi_{\mathrm{d}}$. In addition, we obtain from \eqref{ep} $\left(\frac{d\varphi_{\mathrm{d}}}{dx}\right)^2=\left(\frac{a_0 x}{\dot{a}}\right)^2\frac{\kappa}{6} \epsilon_{\mathrm{d}}$ and from \eqref{Friedmann} $\dot{a}=\sqrt{\frac{\kappa}{3}\epsilon_{\mathrm{d}}}a_0x$, and thus $\left(\frac{d\varphi_{\mathrm{d}}}{dx}\right)^2=\frac{1}{2}$. It then follows that the scalar field describing pure dust comes in two branches $\varphi_{\mathrm{d}}^\pm$ which are for all $x\in\mathds{R}$ exactly given by
\be\lb{fielddust}
\varphi_{\mathrm{d}}^\pm(x)=\mp\frac{x}{\sqrt{2}}+\varphi_{\mathrm{d}}^\pm(0)\ . 
\ee
Here $ \varphi_{\mathrm{d}}^\pm(0)$ denote arbitrary initial values at the present epoch which we set in the following equal to zero without loss of generality. Combinig \eqref{potentialdust} and \eqref{fielddust}, we conclude that the dust model is governed by an exponential potential which reads for the two branches $ \varphi_{\mathrm{d}}^\pm$
\be\lb{potentialphidust}
\tilde{V}_{\mathrm{d}}^\pm\left(\varphi_{\mathrm{d}}^\pm\right)=\frac{\epsilon_0}{2}\mathrm{e}^{\pm 3\sqrt{2}  \varphi_{\mathrm{d}}^\pm(x)}\ . 
\ee
We observe that \eqref{potentialphidust} agrees (for the branch $\varphi_{\mathrm{d}}^+$)with the potential $\tilde{V}_{\mathrm{bb}}$ of Section \ref{sec:bb} used there as an approximation to the potential \eqref{VCH} for the Chaplygin gas close to the big bang and shown in Fig.~\ref{potential} as the dashed curve. (To obtain this agreement, we have to replace the parameter $A$ of the Chaplygin gas by $4\epsilon_0^2$.) There are, however, important differences between the potentials  $\tilde{V}_{\mathrm{bb}}$ and $\tilde{V}_{\mathrm{d}}^+$: i) whereas $\tilde{V}_{\mathrm{bb}}$ is only an \textit{approximation} to the exact potential~\eqref{VCH} in the large-field limit ($\varphi\rightarrow\infty$), the potential $\tilde{V}_{\mathrm{d}}^+$ is \textit{exact} for dust, ii) in the case of the Chaplygin gas, the infinite future of the universe ($x\rightarrow\infty$) is reached for $\varphi\rightarrow 0$, while for the dust component one requires $\varphi_{\mathrm{d}}^\pm\rightarrow\mp\infty$ for $x\rightarrow\infty$. This is obvious by rewriting the energy density~\eqref{energydust} in terms of $\varphi_{\mathrm{d}}^\pm$
\be\lb{energyphidust}
\epsilon_{\mathrm{d}}\left(\varphi_{\mathrm{d}}^\pm\right)=\epsilon_0\mathrm{e}^{\pm 3\sqrt{2}  \varphi_{\mathrm{d}}^\pm(x)}\ . 
\ee
Thus the physically relevant domain in minisuperspace is $(x,\varphi_{\mathrm{d}}^\pm)\in\mathds{R}\times\mathds{R}$ for dust in contrast to $(x,\varphi^\pm)\in\mathds{R}\times\mathds{R}_{\gtrless}$ for the Chaplygin gas.

From \eqref{WDW} and \eqref{potentialphidust} we conclude that the wave function $\Psi_{\mathrm{d}}\left(x,\varphi\right)$ for a universe filled with dust obeys the exact Wheeler-DeWitt equation
\bea\lb{WDWdust}
\hbar^2\left(\frac{\partial^2}{\partial x^2}-\frac{\partial^2}{\partial\varphi_{\mathrm{d}}^2}\right)\Psi_{\mathrm{d}}\left(x,\varphi_{\mathrm{d}}\right)\nonumber\\
+\frac{9}{4}\lambda_{\mathrm{d}}^2 \mathrm{e}^{3\left(2x+\sqrt{2}\varphi_{\mathrm{d}}\right)}\Psi_{\mathrm{d}}\left(x,\varphi_{\mathrm{d}}\right)=0
\ ,  
\eea
where the constant $\lambda_{\mathrm{d}}$ is defined by
\be\lb{lambdadust}
\lambda_{\mathrm{d}}:=2\sqrt{\frac{2\epsilon_0}{3\kappa}}\mathcal{V}_0a_0^3=\frac{2\sqrt{2}H_0}{\kappa}  \mathcal{V}_0a_0^3\ . 
\ee
(Equation~\eqref{WDWdust} is written for the branch $\varphi_{\mathrm{d}}^+$ using the potential $\tilde{V}_{\mathrm{d}}^+$ of Eq.~\eqref{potentialphidust}, and the superscript $+$ is suppressed to simplify the notation.) A comparison with the Wheeler-DeWitt equation~\eqref{WDW*} reveals that \eqref{WDWdust} can be recast into the form
\be\lb{WDWdust*}
\hbar^2\left(\frac{\partial^2}{\partial u^2}-\frac{\partial^2}{\partial v^2}\right)\Psi_{\mathrm{d}}\left(u,v\right)+\Psi_{\mathrm{d}}\left(u,v\right)=0
\ee
using the new coordinates $(u,v)$ which are defined as in equations \eqref{u}-\eqref{nc}, but replacing $\lambda$ by $\lambda_{\mathrm{d}}$ given in \eqref{lambdadust}. Although the range of $(x,\varphi)$ is $\mathds{R}\times\mathds{R}$ in \eqref{WDWdust} instead of ($\mathds{R}\times\mathds{R}_>$) in \eqref{WDW*}, it is obvious that the physically allowed domain of $(u,v)$ for dust is again given by the sector $\mathcal{S}:=\{(u,v)\in\mathds{R}_>\times\mathds{R}, |v|\leq u \}$ indicated in Fig.~\ref{Fig:UVPaper}.

It follows that the most general exact solution to the Wheeler-DeWitt equation~\eqref{WDWdust*} for dust satisfying at the classical big bang singularity DeWitt's criterium \eqref{cond1} respectively \eqref{cond1*} is given by (see equation~\eqref{superposi*})
\be\lb{Psidust}
\begin{split}
\Psi_{\mathrm{d}}(u,v)&=\int_{0}^{\infty}\mathcal{C}_{\mathrm{d}}^+(k) \left[\mathrm{e}^{i\frac{S_+}{\hbar}}-\mathrm{e}^{-i\frac{S_+}{\hbar}}\right]\frac{dk}{\sqrt{2\pi\hbar}}\\&+\int_{0}^{\infty}\mathcal{C}_{\mathrm{d}}^-(k) \left[\mathrm{e}^{i\frac{S_-}{\hbar}}-\mathrm{e}^{-i\frac{S_-}{\hbar}}\right]\frac{dk}{\sqrt{2\pi\hbar}}\ .
\end{split}
\ee
Here the classical actions $S_\pm$ are defined in \eqref{actionbb} with $k\geq 0$ and $\omega(k):=+\sqrt{k^2+1}$. $\mathcal{C}_{\mathrm{d}}^\pm(k)$ denote arbitrary complex "wave number amplitudes". Although the wave packet~\eqref{Psidust} has the form of a WKB-\textit{ansatz}, it is important to remark that \eqref{Psidust} is in the case of dust the \textit{exact} solution to the Wheeler-DeWitt equation~\eqref{WDWdust*}. Furthermore, in contrast to the solution~\eqref{superposi*} for the Chaplygin gas, which is only an approximation close to the big bang, the wave packet \eqref{Psidust} is for a universe filled with dust valid for all cosmic scale factors $a\geq 0$. Since $\Psi_{\mathrm{d}}(u,v)$ in \eqref{Psidust} vanishes for $a\rightarrow0$, that is if one approaches the classical big bang singularity, the latter is avoided in the sense of DeWitt's criterium. It remains to investigate the wave packet \eqref{Psidust} in the semiclassical limit in order to see, whether the amplitudes $\mathcal{C}_{\mathrm{d}}^\pm(k)$ can be chosen in such a way that the wave packet is for $\hbar\rightarrow0$ strongly peaked at the positive branch $\varphi_{\mathrm{d}}^+(x)$ of the scalar field.

Evaluating the wave packet~\eqref{Psidust} in the semiclassical limit using the method of stationary phase (assuming $\mathcal{C}_{\mathrm{d}}^\pm(k)$ to be smooth functions with compact support), one obtains up to corrections of order $\hbar$ that the two integrals in \eqref{Psidust} are proportional to $\mathcal{C}_{\mathrm{d}}^+(k^+_\gamma(u,v))$ respectively $\mathcal{C}_{\mathrm{d}}^-(k^-_\gamma(u,v))$ (see the discussion in Section \ref{semisteiner}), where $k^\pm_\gamma=k^\pm_\gamma(u,v)\in \mathrm{supp}\,\mathcal{C}_{\mathrm{d}}^\pm(k)$ denote the stationary points determined by
\be\lb{stationaryphasedust}
\begin{split}
\frac{\partial S^{\pm}\left(k_\gamma^\pm(u,v);u,v\right)}{\partial k}=\frac{k_\gamma^\pm(u,v)}{\left[\left(k_\gamma^\pm(u,v)\right)^2+1\right]^\frac{1}{2}}&=0 \ ,\\
\frac{\partial^2 S^{\pm}\left(k_\gamma^\pm(u,v);u,v\right)}{\partial k^2}=\frac{u}{\left[\left(k_\gamma^\pm(u,v)\right)^2+1\right]^\frac{3}{2}}&\not=0 \ .
\end{split}
\ee
The only stationary points are then given by
\be\lb{kdust}
k_\gamma^\pm(u,v)=\mp \frac{v}{\sqrt{u^2-v^2}} \ , 
\ee
where $k^\pm_\gamma$ corresponds to $v\lessgtr 0$, keeping in mind that $k>0$ must hold. Furthermore, the second derivative in \eqref{stationaryphasedust} takes the value $u\left[1-\left(\frac{v}{u}\right)^2\right]^{\frac{3}{2}}$ which is nonzero in the interior of the physical region, i.e. for $0<|v|<u$.

Let us now assume that both amplitudes $\mathcal{C}_{\mathrm{d}}^\pm(k)$ are strongly peaked at the classical wave number
\be\lb{kcl}
k_{\mathrm{cl}}:=1 \ . 
\ee
It then follows that the wave packet $\Psi_{\mathrm{d}}(u,v)$ is in the semiclassical limit strongly peaked at $ v^\pm:=\mp\frac{u^\pm}{\sqrt{2}}$ which in turn implies that it is strongly peaked precisely at the classical field configurations $\varphi_{\mathrm{d}}^\pm (x)= \mp \frac{x}{\sqrt{2}}$, in agreement with the exact solutions~\eqref{fielddust} (for $\varphi_{\mathrm{d}}^\pm (0)=0 $). The last statement is easily shown to be true, e.g. for $\varphi_{\mathrm{d}}^+ (x)$, by calculating $u^+:=u(x,\varphi_{\mathrm{d}}^+)=(\sigma+\rho)\frac{\lambda_{\mathrm{d}}}{6} \mathrm{e}^{\frac{\sigma\rho}{3}x}$ and $v^+:=v(x,\varphi_{\mathrm{d}}^+)=-(\sigma-\rho)\frac{\lambda_{\mathrm{d}}}{2} \mathrm{e}^{\frac{\sigma\rho}{3}x}$ and noting that $\frac{\sigma-\rho}{\sigma+\rho}=\frac{1}{\sqrt{2}}$, which implies the correct relation $v^+=-\frac{u^+}{\sqrt{2}}$.

It is now obvious that one can fulfill the remaining boundary condition which guarantees that the wave packet \eqref{Psidust} is in the semiclassical limit only peaked at the positive branch of the scalar field (corresponding to the choice of the potential $\tilde{V}_{\mathrm{d}}^+ (\varphi_{\mathrm{d}}^+)$ considered in the Wheeler-DeWitt equation~\eqref{WDWdust}) by imposing the condition $\mathcal{C}_{\mathrm{d}}^-(k)\equiv 0$.

It has thus been demonstrated that in the case of a universe filled with dust there exists a quantum wave packet derived from the exact Wheeler-DeWitt equation which avoids the classical big bang singularity and is strongly peaked for $\hbar\rightarrow 0$ precisely at the exact classical field $\varphi_{\mathrm{d}}^+$, valid at all cosmic scale factors $a>0$.
\section{Summary and Discussion}\lb{discussion}
In this paper, we have studied a Friedmann--Lema\^{\i}tre universe whose energy/matter content is described by a scalar field $\phi$ obeying either the equation of state of a Chaplygin gas or that of pure dust. In Sec.~\ref{mini} the general symmetry reduction to homogeneous and isotropic geometries has been performed in the frame work of classical cosmology starting from the Einstein-Hilbert action and the action of a scalar field minimally coupled to gravity and governed by an arbitrary potential $V(\phi)$. The canonical Hamiltonian of the two-dimensional minisuperspace $(a,\phi)$ obtained this way is constrained to vanish and leads to the standard Friedmann equation for the cosmic scale factor $a(t)$ and to the Klein-Gordon equation for $\phi(t)$. Specializing to the case of a Chaplygin gas completely determines the scalar potential $V(\phi)$, see equation~\eqref{VCH}. The classical time evolution of $a(t)$ and $\phi(t)$ has been investigated in detail in Sec.~\ref{klass}, and it is seen that the universe starts in this model with a classical big bang as in a pure dust model and undergoes at late times an accelerated expansion dominated by a dark enery component as in models with a positive cosmological constant. The results of this model have been illustrated in Figs.~\ref{figDENSITY}-\ref{Fig:UVPaper} for a particular value of the parameter $A$ (see equation~\eqref{state}), being the only free parameter characterizing the Chaplygin gas, which is chosen in such a way that the age of the universe is equal to the Hubble time.

In Sec.~\ref{sec:quant} the quantization of this model has been carried out in the framework of geometrodynamics leading to the Wheeler-DeWitt equation for the "wave function of the universe" $\Psi$. For a general scalar field, the Wheeler-DeWitt equation in the two-dimensional minisuperspace choosing the Laplace-Beltrami factor ordering has been given in equation~\eqref{WDW} using the dimensionless variables $x:=\ln\,\left(\frac{a}{a_0}\right)$ and $\varphi:=\sqrt{\frac{\kappa}{6}}\phi$. Since the exact solution to the Wheeler-DeWitt equation valid in the whole minisuperspace is not known to us, we have studied in Sec.~\ref{sec:bb} the approximation to the Wheeler-DeWitt equation which holds close to the classical big bang singularity (corresponding to $|\varphi|\rightarrow\infty$), and in Sec.~\ref{sec:largevalues} the alternative approximation valid for large scale factors respectively small scalar fields $\varphi\rightarrow 0$. In both limits, the complete solutions to the Wheeler-DeWitt equation have been found, see \eqref{Psi0} respectively equation~\eqref{Psiunend}. These solutions depend on arbitrary coefficients which have to be fixed by boundary conditions as one is familiar from standard quantum mechanics. 

In Sec.~\ref{BB} we have shown that the classical big bang singularity can be avoided if one imposes DeWitt's initial condition \eqref{cond1} ascertaining that the wave function vanishes on the hypersurface $a=0$ for any given scalar field $\varphi$. Under this condition, the remaining freedom in the wave function is already drastically reduced as shown in equation~\eqref{Psi0*}. To reduce this freedom even further, we have proposed to supplement DeWitt's boundary condition at the big bang by a second boundary condition to be imposed in the semiclassical limit in which Planck's constant approaches zero. Since in this limit the standard classical cosmology should be recovered, we demand that the wave packet constructed from the solution~\eqref{Psi0*} should be strongly peaked (with a suitable choice of the wave number amplitude) at the correct classical solution of the scalar field $\varphi$. In this way we then obtained the final solution~\eqref{superposi**} of the Wheeler-DeWitt equation which has the form of a WKB-wave function even though it solves the Wheeler-DeWitt equation exactly. The fact that the wave packet~\eqref{superposi**} depends only on the classical action $S_+$ (corresponding to the classical outgoing field) automatically guarantees that our second condition is fulfilled in the semiclassical limit as we have shown by the method of stationary phase.

In Sec.~\ref{semisteiner} we have explained in detail how the method of stationary phase is applied in the semiclassical limit to the wave function~\eqref{Psiunend} which represents the complete solution to the Wheeler-DeWitt equation~\eqref{WDW***} in the case of large values of the scale factor, that is for the universe in the far future. Employing the Debye type asymptotics of the Hankel functions, we have constructed the general wave packet~\eqref{PsiStein*} which gives the dominant term in an asymptotic expansion for $\hbar\rightarrow 0$. Under the assumption that the wave number amplitudes $\mathcal{B}_i(K)$ are smooth functions with compact support and are strongly peaked at the classical wave number $K_{\mathrm{cl}}$ (given in Eq.~\eqref{K_cl}), we have shown that the wave packet is strongly peaked at the semiclassical field trajectories $\varphi_{\mathrm{scl}}^\pm(x)$ \eqref{phicl} respectively \eqref{phicl-}. The main point to observe is that these semiclassical fields possess for $a\gg a_0$ in leading approximation exactly the same behavior as the exact classical fields $\varphi_{\mathrm{cl}}^\pm(x)$ given in \eqref{phiclpm}.

Finally, in Sec.~\ref{dust} we have studied the quantum cosmology of a pure dust model which is governed by the exponential potentials~\eqref{potentialphidust} corresponding to the two branches of the classical field. In this case the Wheeler-DeWitt equation~\eqref{WDWdust} holds in the whole minisuperspace without any restriction. Imposing DeWitt's initial condition~\eqref{cond1}, we have derived the general wave packet~\eqref{Psidust} satisfying the Wheeler-DeWitt equation and have thus found the exact quantum wave packet for a universe filled with dust which avoids the classical big bang singularity. Taking into account the second boundary condition in the semiclassical limit, we have shown that from~\eqref{Psidust} one can construct a wave packet which is strongly peaked e.g. at the \textit{exact} field $\varphi_{\mathrm{d}}^+(x)$ corresponding to the dust potential $\tilde{V}^+_{\mathrm{d}}$ in \eqref{potentialphidust}.

We would like to remark that the various solutions to the Wheeler-DeWitt equation have been obtained under the assumption that this equation can be treated as in ordinary quantum mechanics. The question whether the Wheeler-DeWitt equation corresponds to a self adjoint operator in a Hilbert space equipped with a scalar product leading to a consistent probability interpretation has not been dicussed in this paper. To our knowledge, there exists up to now no complete mathematical treatment concerning these problems in the framework of geometrodynamics (see however \cite{text,Craig:10}). It is therefore of great importance that there exists an alternative approach toward the quantization of gravity, namely LQG respectively LQC if applied to cosmology. In this theory the avoidance of singularities could be accomplished according to \cite{Thiemann_07,Rovelli_04,Ashtekar_04,BojowaldR}. Moreover, the LQC \textit{ansatz} can be used to make physical predictions in the vicinity of the big bang. As shown in \cite{Woehr:09} the situation is improved with respect to the singularity avoidance in scalar field models such as the (anti-)Chaplygin gas, inflationary, radiation dominated or quintessence models. \\

\section*{ACKNOWLEDGMENTS}
A.W. would like to thank Raphael Lamon for fruitful discussions and comments and Holger Stefan Janzer for technical support.



\begin{thebibliography}{99}
\bibitem{DeWitt_67}  
B. S. DeWitt, {\em Quantum Theory of Gravity I: The Canonical Theory }. Phys. Rev. {\bf 160}, 1113 (1967).
\bibitem{Misner_69}  
C. W. Misner, {\em Quantum Cosmology. I}. Phys. Rev. {\bf 186}, 1319 (1969).
\bibitem{C.Kiefer_07}
C. Kiefer, {\em Quantum Gravity}. Second Edition (Oxford University
Press, Oxford, 2007).
\bibitem{Thiemann_07}
T. Thiemann,
{\em Modern Canonical Quantum General Relativity}.
(Cambridge University Press, Cambridge, 2007)   
\bibitem{BojowaldR} 
M. Bojowald, Living Rev. Relativity {\bf 8}, 11 (2005).
\bibitem{Ashtekar_04}  
A. Ashtekar and L. Lewandowski,{\em Background independent quantum gravity: A status report}. Class. Quant. Grav. {\bf 21}, R53-R152 (2004).
\bibitem{Thiemann_03}
T. Thiemann, {\em Lectures on Loop Quantum Gravity}. Lect. Notes. Phys. {\bf631}, 41 (2003).
\bibitem{Rovelli_04}
C. Rovelli, {\em Quantum Gravity}.(Cambridge University
Press, Cambridge, 2004).
\bibitem{Ashtekar_06*}  
A. Ashtekar,
  T. Pawlowski, and P. Singh, Phys. Rev. D{\bf 74}, 084003 (2006).
\bibitem{Ashtekar_07} 
A. Ashtekar, T. Pawlowski, P. Singh, and
  K. Vandersloot, Phys. Rev. D {\bf 75}, 024035 (2007). 
\bibitem{Wheeler:64}
J. A. Wheeler, {\em Geometrodynamics}. (Academic Press, New York, 1964).
\bibitem{Ashtekar_06} 
A. Ashtekar, M. Bojowald, and J. Lewandowski,
  Adv. Theor. Math. Phys. {\bf 7}, 233 (2003); A. Ashtekar,
  T. Pawlowski, and P. Singh, Phys. Rev. D{\bf 73}, 124038 (2006); A. Ashtekar,
  T. Pawlowski, and P. Singh, Phys. Rev. Lett. {\bf 96}, 141301 (2006). 
\bibitem{Chap_04}
S. Chaplygin, Sci. Mem. Moskov Univ. Math. Phys. {\bf 21}, 1 (1904).
\bibitem{KAM_01}
A. Kamenshchik, U. Moschalla, and V. Pasquir, {\em An alternative to quintessence}.
 Phys. Lett. {\bf{B 511}}, 265 (2001).  
\bibitem{Fabris_02}
J. C. Fabris, S. V. B. Goncalves and P. E. de Souza, {\em Density perturbations in a Universe
dominated by the Chaplygin gas}. Gen. Rel. Grav. {\bf 34}, 1 (2002).
\bibitem{Fabris_02*}
J. C. Fabris, S. V. B. Goncalves and P. E. de Souza, {\em Mass power spectrum in a Universe
dominated by the Chaplygin gas}. Gen. Rel. Grav. {\bf 34}, 12 (2002)
\bibitem{Gorini_03}
V. Gorini, A. Kamenshchik and U. Moschella, {\em Can the Chaplygin gas be a plausible model for
Dark Energy?} Phys. Rev. D {\bf 67}, 063509 (2003)
\bibitem{Gorini_05}
V. Gorini, A. Kamenshchik, U. Moschella and V. Pasquier, {\em The Chaplygin gas as a model for
Dark Energy}. In: Proc. MG10, Rio de Janeiro, Brazil, 20-26 July 2003, M. Novello, S. Perez
Bergliaffa, R. Ruffini (eds.). World Scientific, p. 840 (2005).
\bibitem{Chimento_03}
L. P. Chimento, A. S. Jakubi and D. Pavo, {\em Dark Energy, dissipation, and the coincidence
problem}. Phys. Rev.  D {\bf 67}, 087302 (2003).
\bibitem{Chimento_04}
L. P. Chimento, {\em Extended tachyon field, Chaplygin gas, and solvable k-essence cosmologies}.
Phys. Rev. D {\bf 69}, 123517 (2004).
\bibitem{Dev_03}
A. Dev, J. S. Alcaniz and D. Jain, {\em Cosmological consequences of a Chaplygin gas Dark Energy}.
Phys. Rev. D {\bf 67}, 023515  (2003).
\bibitem{Avelino_03}
P. P. Avelino, L. M. G. Beca, J. P.M. de Carvalho, C. J. A. P. Martins and P. Pinto, {\em  Alternatives
to quintessence model building}. Phys. Rev. D {\bf 67}, 023511 (2003).
\bibitem{Bean_03}
R. Bean and O. Dore, {\em Are Chaplygin gases serious contenders for the Dark Energy?} Phys. Rev.
D {\bf 68}, 023515 (2003).
\bibitem{Carturan_03}
D. Carturan and F. Finelli, {\em Cosmological effects of a class of fluid Dark Energy models}. Phys.
Rev. D {\bf 68}, 103501 (2003).
\bibitem{Bordemann_93}
M. Bordemann and J. Hoppe, {\em The dynamics of relativistic membranes. Reduction to 2-
dimensional fluid dynamics}. Phys. Lett. B {\bf317}, 315 (1993).
\bibitem{Jackiw_00}
R. Jackiw and A. P. Polychronakos, {\em Supersymmetric fluid mechanics}. Phys. Rev. {\bf D} 62, 085019
(2000).
\bibitem{KAM_07}
A. Kamenshchik, C. Kiefer, and B. Sandh\"ofer, {\em Quantum cosmology with a big-brake singularity}.
Phys. Rev. {\bf{D 76}}, 064032 (2007).
\bibitem{MAGNUS_66}
W. Magnus, F. Oberhettinger, and R.P. Soni,
{\em Formulas and Theorems for the Special Functions of Mathematical Physics}.
Third edition	(Springer-Verlag Berlin Heidelberg New York, 1966). 
\bibitem{Grigis:94}
A. Grigis and J. Sj\"ostrand, {\em Microlocal Analysis for Differential Operators}.
London Math. Soc. Lecture Note Series {\bf 196}. (Cambridge University Press, Cambridge, 1994).
\bibitem{text}
After completion of this paper, we became aware of the recent preprint \cite{Craig:10}.
\bibitem{Craig:10}
D. A. Craig, and P. Singh, {\em Consistent Histories in Quantum Cosmology}.
\texttt{arXiv:1001.4311} (2010).
\bibitem{Woehr:09}
R. Lamon, and A. J. W\"ohr, {\em Quintessence and (anti-)Chaplygin Gas in Loop Quantum Cosmology}.
Phys. Rev. D {\bf 81}, 024026 (2010).
\end{thebibliography}
\end{document}